\documentstyle[12pt,epsf]{article}

\def\spose#1{\hbox to 0pt{#1\hss}}
\def\lsim{\mathrel{\spose{\lower 3pt\hbox{$\mathchar"218$}}
 \raise 2.0pt\hbox{$\mathchar"13C$}}}
\def\gsim{\mathrel{\spose{\lower 3pt\hbox{$\mathchar"218$}}
 \raise 2.0pt\hbox{$\mathchar"13E$}}}

\begin{document}

\begin{titlepage}

\begin{flushright}
CERN-TH/97-240\\
hep-ph/9709327
\end{flushright}

\vspace{3cm}
\begin{center}
\boldmath
\Large\bf
Theoretical Analysis of $\bar B\to D^{**}\pi$ Decays
\unboldmath
\end{center}

\vspace{0.5cm}
\begin{center}
M. Neubert\\[0.1cm]
{\sl Theory Division, CERN, CH-1211 Geneva 23, Switzerland}
\end{center}

\vspace{1.5cm}
\begin{abstract}
\vspace{0.2cm}\noindent
The decays $\bar B\to D^{**}\pi$, where $D^{**}$ is a narrow p-wave
charm resonance, are investigated in the context of a generalized
factorization hypothesis, including the leading nonfactorizable
corrections. The decay rates for these processes are related to the
corresponding semileptonic rates at maximum recoil. It is pointed out
that the current data on the branching ratio for the decay $B^-\to
D_2^{*0}\,\pi^-$ may pose a problem for theory. We predict that future,
more accurate measurements of this branching ratio will find a value
${\mathrm B}(B^-\to D_2^{*0}\,\pi^-)\approx 4\times 10^{-4}$, a factor
5 lower than the current central value.
\end{abstract}

\vfill
\noindent
CERN-TH/97-240\\
September 1997

\end{titlepage}

\section{Introduction}

Recently, the CLEO Collaboration has reported the first measurement of
the branching ratios for the decays $B^-\to D_1^0\,\pi^-$ and $B^-\to
D_2^{*0}\,\pi^-$, where $D_1$ and $D_2^*$ (generically denoted as
$D^{**}$) are p-wave charm mesons. In the heavy-quark limit, these are
the members of a spin-symmetry doublet in which the light degrees of
freedom have total angular momentum $j=\frac 32$. The measured
branching ratios are \cite{CLEOhad}
\begin{eqnarray}
   {\mathrm B}(B^-\to D_1^0\,\pi^-)\times
   {\mathrm B}(D_1^0\to D^{*+}\pi^-) &=& (7.8\pm 1.9)\times 10^{-4}
    \,, \nonumber\\
   {\mathrm B}(B^-\to D_2^{*0}\,\pi^-)\times
   {\mathrm B}(D_2^{*0}\to D^{*+}\pi^-) &=& (4.2\pm 1.7)\times 10^{-4}
    \,.
\label{CLdat1}
\end{eqnarray}
The purpose of this letter is to interpret these observations together
with the existing data on the semileptonic decays $B^-\to
D_1^0\,\ell^-\bar\nu$ and $B^-\to D_2^{*0}\,\ell^-\bar\nu$. For the
corresponding branching ratios, the CLEO Collaboration quotes
\cite{CLEOsl}
\begin{eqnarray}
   {\mathrm B}(B^-\to D_1^0\,\ell^-\bar\nu)\times
   {\mathrm B}(D_1^0\to D^{*+}\pi^-) &=& (0.37\pm 0.10)\% \,,
    \nonumber\\
   {\mathrm B}(B^-\to D_2^{*0}\,\ell^-\bar\nu)\times
   {\mathrm B}(D_2^{*0}\to D^{*+}\pi^-) &=& (0.059\pm 0.067)\%
    \nonumber\\
   &<& 0.16\% \quad (90\%~{\mathrm CL}) \,.
\label{CLdat2}
\end{eqnarray}
These numbers are consistent with results reported by the ALEPH
Collaboration \cite{ALEPHsl}:
\begin{eqnarray}
   {\mathrm B}(\bar B\to D_1\,\ell^-\bar\nu X)\times
   {\mathrm B}(D_1\to D^*\pi^\pm) &=& (0.49\pm 0.11)\% \,,
    \nonumber\\
   {\mathrm B}(\bar B\to D_2^{*0}\,\ell^-\bar\nu X)\times
   {\mathrm B}(D_2^{*0}\to D^{*+}\pi^-) &<& 0.34\%
    \quad (95\%~{\mathrm CL}) \,, \nonumber\\
   {\mathrm B}(\bar B\to D_2^{*0}\,\ell^-\bar\nu X)\times
   {\mathrm B}(D_2^{*0}\to D^+\pi^-) &<& 0.33\%
    \quad (95\%~{\mathrm CL}) \,, \nonumber\\
   {\mathrm B}(\bar B\to D_2^{*+}\,\ell^-\bar\nu X)\times
   {\mathrm B}(D_2^{*+}\to D^0\,\pi^+) &<& 0.26\%
    \quad (95\%~{\mathrm CL}) \,,
\end{eqnarray}
if one assumes that in most cases no additional particles $X$ are
produced. Semileptonic $B$ decays into p-wave charm mesons are
important with regard to understanding how the total inclusive
semileptonic branching ratio of $B$ mesons is composed out of exclusive
modes. The ALEPH Collaboration finds a sizable branching ratio of
$(2.26\pm 0.44)\%$ for the sum of all (resonant or nonresonant) decays
$\bar B\to D^{(*)}\pi\,\ell^-\bar\nu$ \cite{ALEPHsl}. These channels
make up for about 20\% of the total semileptonic branching ratio.

Unfortunately, the $D^{**}\to D^{(*)}\pi$ branching ratios entering the
above measurements have not yet been determined experimentally. Under
the assumption that all $D_1$ mesons decay into $D^*\pi$ (decays into
$D\,\pi$ are forbidden by angular momentum conservation), and using
isospin invariance, one obtains ${\mathrm B}(D_1^0\to
D^{*+}\pi^-)=2/3$. For $D_2^*$ mesons, the ratio ${\mathrm
B}(D_2^{*0}\to D^+\pi^-)/{\mathrm B}(D_2^{*0}\to D^{*+}\pi^-)=2.3\pm
0.8$ has been measured \cite{CLE,ARG}. Under the assumption that
$D_2^*\to D\,\pi$ and $D_2^*\to D^*\pi$ are the only decay modes, and
using isospin invariance, one then obtains ${\mathrm B}(D_2^{*0}\to
D^{*+}\pi^-)=(20\pm 5)\%$ and ${\mathrm B}(D_2^{*0}\to
D^+\pi^-)={\mathrm B}(D_2^{*+}\to D^0\,\pi^+)= (47\pm 5)\%$.

In the case of neutral $B$ mesons, there is a simple relation between
hadronic decays involving a negatively charged pion in the final state
and the corresponding semileptonic decays with the pion replaced by a
lepton--neutrino pair \cite{Bj89}. In general, the amplitude for the
hadronic decay $\bar B^0\to H_c^+\pi^-$, where $H_c^+$ represents an
arbitrary charm meson (or a sum of hadrons with total charm number
one), can be written as \cite{BSW}
\begin{equation}
   A_{\mathrm had} = \frac{G_F}{\sqrt 2}\,V_{cb} V_{ud}^*\,a_1\,
   \langle\pi^-|\,\bar d\gamma_\mu(1-\gamma_5)u\,|\,0\,\rangle\,
   \langle H_c^+|\,\bar c\gamma^\mu(1-\gamma_5)b\,|\bar B^0\rangle \,,
\label{class1}
\end{equation}
where $a_1$ parametrizes the complicated hadronic matrix elements of
four-quark operators, which enter the theoretical description of
nonleptonic amplitudes. This will be discussed in more detail below.
The squared amplitude takes the form
\begin{equation}
   |A_{\mathrm had}|^2 = \frac{G_F^2}{2}\,|V_{cb} V_{ud}^*|^2\,
   |a_1|^2 f_\pi^2\,q_\mu q_\nu H^{\mu\nu} \,,
\end{equation}
where $q=p-p'$ is the pion momentum, $f_\pi$ the pion decay constant,
and
\begin{equation}
   H^{\mu\nu} = \langle H_c^+|\,\bar c\gamma^\mu(1-\gamma_5)b\,
   |\bar B^0\rangle\,\langle \bar B^0|\,
   \bar b\gamma^\nu(1-\gamma_5)c\,|H_c^+\rangle
\end{equation}
the ``hadronic tensor''. The same tensor appears in the description of
semileptonic decays. Indeed, the squared amplitude for the process
$\bar B^0\to H_c^+\,\ell^-\bar\nu$ is given by
\begin{equation}
   |A_{\mathrm sl}|^2 = \frac{G_F^2}{2}\,|V_{cb}|^2
   L_{\mu\nu} H^{\mu\nu} \,,
\end{equation}
where $L_{\mu\nu}$ is the leptonic tensor. In the special case where
$q^2=(p_\ell+p_\nu)^2=0$, the leptonic tensor takes the simple form
\begin{equation}
   L_{\mu\nu}\Big|_{q^2=0} = 16 x(1-x) q_\mu q_\nu \,,
\end{equation}
where $x=E_\ell/E_\ell^{\mathrm max}$ is the scaled lepton energy in
the rest frame of the $B$ meson. Thus, in the limit where the pion mass
is neglected, the nonleptonic decay rate is proportional to the
semileptonic rate at $q^2=0$. Working out the trivial phase-space
factors, we obtain Bjorken's relation \cite{Bj89}
\begin{equation}
   \frac{\Gamma(\bar B^0\to H_c^+\pi^-)}
   {{\mathrm d}\Gamma(\bar B^0\to H_c^+\ell^-\bar\nu)/
    {\mathrm d}q^2 \Big|_{q^2=0}}
   = 6\pi^2 f_\pi^2\,|V_{ud}|^2\,|a_1|^2
   + O\left( \frac{M_\pi^2}{M_B^2} \right)
\label{ratio}
\end{equation}
irrespective of the nature of the hadron state $H_c$. The corrections
of order $M_\pi^2/M_B^2$ have been worked out for the cases $H_c=D$ and
$D^*$ and are found to be negligible \cite{NS}. Relation (\ref{ratio})
still contains an unknown hadronic parameter $a_1$. However, for
energetic two-body decays such as $\bar B^0\to D^{(*)+}\pi^-$ one can
argue that the value of this parameter is close to unity. The reason is
that the pion has a large energy in the rest frame of the decaying $B$
meson. Once its constituent quarks have grouped together in a
colour-singlet state, they form a fast-moving colour dipole which
decouples from long-wavelength gluons. Only hard gluons with
virtualities of order $M_B$ are effective in rearranging the quarks.
This ``colour transparency argument'' \cite{Bj89} suggests that the
nonfactorizable contributions to the hadronic decay amplitude are
switched off at low momentum scales. Hence, one expects that
$a_1\approx 1$ for energetic two-body decays. This assertion is further
strengthened by the $1/N_c$ expansion, which shows that
$a_1=1+O(1/N_c^2)$ \cite{NS}. A determination of $a_1$ from $\bar
B^0\to D^{(*)+}\pi^-$ decays gives $|a_1|=1.08\pm 0.11$, supporting the
theoretical arguments just presented. A similar value is expected to
apply for two-body decays into p-wave charm mesons.

Unfortunately, the situation is more complicated for the decays of
charged $B$ mesons. Whereas the semileptonic decay rates of $B^-$ and
$\bar B^0$ mesons are related to each other by isospin invariance, this
is not so for the hadronic rates. We will now discuss how relation
(\ref{ratio}) must be modified in this case.

\boldmath
\section{Hadronic decay amplitudes for $\bar B\to D^{**}\pi$}
\unboldmath

The part of the effective weak Hamiltonian relevant to $b\to c\bar u d$
transitions is given by
\begin{equation}
   H_{\mathrm eff} = \frac{G_F}{\sqrt 2}\,V_{cb} V_{ud}^*\,
   \Big\{ c_1(\mu)\,(\bar d u) (\bar c b)
   + c_2(\mu)\,(\bar c u) (\bar d b) + \dots \Big\} \,,
\end{equation}
where $(\bar d u)=\bar d\gamma^\mu(1-\gamma_5) u$ etc.\ are
left-handed, colour-singlet quark currents. The Wilson coefficients
$c_i(\mu)$ are known to next-to-leading order. At the scale $\mu=m_b$,
they have the values $c_1(m_b)\approx 1.1$ and $c_2(m_b)\approx -0.3$.
These coefficients take into account the short-distance corrections
arising from the exchange of hard gluons. The effects of soft gluons
(with virtualities below the scale $\mu$) remain in the hadronic matrix
elements of the local four-quark operators. A reliable field-theoretic
calculation of these matrix elements is the obstacle to a quantitative
theory of hadronic weak decays.

Using Fierz identities, the four-quark operators in the effective
Hamiltonian may be rewritten in various forms. It is particularly
convenient to rearrange them in such a way that the flavour quantum
numbers of one of the quark currents match those of one of the hadrons
in the final state of the considered decay process. In the case of
$\bar B\to D^{**}\pi$ transitions, omitting common factors, the various
decay amplitudes may be written as
\begin{eqnarray}
   A(\bar B^0\to D^{**+}\pi^-) &=& \bigg( c_1 + \frac{c_2}{N_c} \bigg)
    \langle D^{**+}\pi^-|(\bar d u)(\bar c b)|\bar B^0\rangle
    \nonumber\\
   &&\mbox{}+ 2 c_2\,\langle D^{**+}\pi^-|(\bar d t_a u)
    (\bar c t_a b)|\bar B^0\rangle \,, \nonumber\\
   A(\bar B^0\to D^{**0}\pi^0) &=& \bigg( c_2 + \frac{c_1}{N_c} \bigg)
    \langle D^{**0}\pi^0|(\bar c u)(\bar d b)|\bar B^0\rangle
    \nonumber\\
   &&\mbox{}+ 2 c_1\,\langle D^{**0}\pi^0|(\bar c t_a u)
    (\bar d t_a b)|\bar B^0\rangle \,, \nonumber\\
   A(B^-\to D^{**0}\pi^-) &=& A(\bar B^0\to D^{**+}\pi^-)
    - \sqrt 2\,A(\bar B^0\to D^{**0}\pi^0) \,,
\label{ampl}
\end{eqnarray}
where $t_a$ are the SU(3) colour matrices. The last relation follows
from isospin symmetry of the strong interactions. The three classes of
decays shown above are referred to as class-1, class-2, and class-3,
respectively \cite{BSW}.

The class-1 amplitude $A(\bar B^0\to D^{**+}\pi^-)$ contains the
``factorizable contribution''
\begin{equation}
   A_{\mathrm fact} = \langle\pi^-|(\bar d u)|\,0\,\rangle\,
   \langle D^{**+}|(\bar c b)|\bar B^0\rangle \,,
\label{Afact}
\end{equation}
which can be calculated in terms of the pion decay constant $f_\pi$ and
the $\bar B^0\to D^{**+}$ transition form factors. It also contains
other, nonfactorizable contributions, which can be accounted for by
introducing hadronic parameters $\varepsilon_1$ and $\varepsilon_8$
such that
\begin{eqnarray}
   \langle D^{**+}\pi^-|(\bar d u)(\bar c b)|\bar B^0\rangle
   &=& \left[ 1 + \varepsilon_1(\mu) \right]\,A_{\mathrm fact} \,,
    \nonumber\\
   \langle D^{**+}\pi^-|(\bar d t_a u)(\bar c t_a b)|\bar B^0\rangle
   &=& \frac 12 \varepsilon_8(\mu)\,A_{\mathrm fact} \,.
\end{eqnarray}
Then the class-1 decay amplitude takes the form shown in
(\ref{class1}), i.e.\ $A(\bar B^0\to D^{**+}\pi^-) =
a_1\,A_{\mathrm fact}$, with \cite{NS}--\cite{Soar}
\begin{equation}
   a_1 = \bigg( c_1(\mu) + \frac{c_2(\mu)}{N_c} \bigg) \left[ 1
   + \varepsilon_1(\mu) \right] + c_2(\mu)\,\varepsilon_8(\mu) \,.
\end{equation}
The hadronic parameter $a_1$ takes into account all contributions to
the matrix elements and is thus $\mu$ independent. The scale dependence
of the Wilson coefficients is exactly balanced by that of the hadronic
parameters $\varepsilon_i(\mu)$.

In the case of $\bar B\to D^{(*)}\pi$ transitions, a similar discussion
can be done for the class-2 amplitude, which contains the factorizable
contribution $\langle D^{(*)0}|(\bar c u)|\,0\,\rangle\,
\langle\pi^0|(\bar d b)|\bar B^0\rangle$. Nonfactorizable contributions
can be accounted for by introducing a parameter $a_2$, which has a
similar structure as $a_1$ except for an interchange of $c_1$ and
$c_2$. In the present case of $\bar B\to D^{**}\pi$ transitions,
however, things are more subtle, since the factorizable contribution to
the class-2 decay amplitude vanishes. The reason is that the p-wave
charm mesons $D_1$ and $D_2^*$ do not couple to the vector or axial
vector currents, i.e. $\langle D^{**0}|(\bar c u)|\,0\,\rangle=0$.
Nevertheless, the matrix elements of the four-quark operators need not
vanish. Taking into account that there is only a single helicity
amplitude for the decays considered here, we choose to normalize these
matrix elements to the factorized class-1 amplitude $A_{\mathrm fact}$
introduced in (\ref{Afact}) and define
\begin{eqnarray}
   -\sqrt 2\,\langle D^{**0}\pi^0|(\bar c u)(\bar d b)|\bar B^0\rangle
   &=& \delta_1(\mu)\,A_{\mathrm fact} \,, \nonumber\\
   -\sqrt 2\,\langle D^{**0}\pi^0|(\bar c t_a u)(\bar d t_a b)
   |\bar B^0\rangle
   &=& \frac 12 \delta_8(\mu)\,A_{\mathrm fact} \,.
\end{eqnarray}
Then the class-2 amplitude takes the form $-\sqrt 2\,A(\bar B^0\to
D^{**0}\pi^0) = \bar a_2\,A_{\mathrm fact}$, where
\begin{equation}
   \bar a_2 = \bigg( c_2(\mu) + \frac{c_1(\mu)}{N_c} \bigg)
   \delta_1(\mu) + c_1(\mu)\,\delta_8(\mu) \,.
\end{equation}

Additional insight can be gained by combining these results with the
$1/N_c$ expansion \cite{NS}. At a scale $\mu=O(m_b)$, the large-$N_c$
counting rules of QCD imply\footnote{For scales much lower than $m_b$,
the counting rules for the Wilson coefficients $c_i(\mu)$ are spoiled
by large logarithms.}
$c_1=1+O(1/N_c^2)$ and $c_2=O(1/N_c)$ for the Wilson coefficients, and
$\varepsilon_1, \delta_1=O(1/N_c^2)$ and $\varepsilon_8,
\delta_8=O(1/N_c)$ for the hadronic parameters. Using these results, we
find
\begin{equation}
   a_1 = 1 + O\left( \frac{1}{N_c^2} \right) \,, \qquad
   \frac{\bar a_2}{a_1} = \delta_8(m_b) + O\left( \frac{1}{N_c^3}
   \right) \,.
\end{equation}
Hence, for the class-1 amplitude we recover the result $a_1\approx 1$,
which also follows from the colour-transparency argument. The class-2
amplitude, on the other hand, is governed by a nontrivial hadronic
parameter $\delta_8(m_b)$ of order $1/N_c$, which is process depend and
will, in general, take different values for $D_1$ and $D_2^*$. For the
ratios of the various hadronic decay rates, we obtain
\begin{eqnarray}
   \frac{\Gamma(\bar B^0\to D^{**0}\pi^0)}
    {\Gamma(\bar B^0\to D^{**+}\pi^-)}
   &\approx& \frac 12\,|\delta_8(m_b)|^2 \,, \nonumber\\
   \frac{\Gamma(B^-\to D^{**0}\pi^-)}
    {\Gamma(\bar B^0\to D^{**+}\pi^-)}
   &\approx& |1+\delta_8(m_b)|^2 \,.
\end{eqnarray}
Whereas the first ratio may be used to extract the magnitude (but not
the phase) of $\delta_8(m_b)$, the second one determines the
corrections to Bjorken's relation (\ref{ratio}) for the case of class-3
decays:
\begin{equation}
   \frac{\Gamma(B^-\to D^{**0}\pi^-)}
   {{\mathrm d}\Gamma(B^-\to D^{**0}\ell^-\bar\nu)/
    {\mathrm d}q^2 \Big|_{q^2=0}}
   \approx 6\pi^2 f_\pi^2\,|V_{ud}|^2\,|1+\delta_8(m_b)|^2 \,.
\label{cl3ratio}
\end{equation}
Note that the final state $D^{**0}\pi^-$ is a pure isospin $I=\frac 32$
state. Therefore, the class-3 decays are unaffected by (elastic)
final-state interactions.

\boldmath
\section{Semileptonic decay amplitudes for $\bar B\to
D^{**}\ell^-\bar\nu$}
\unboldmath

The theoretical description of semileptonic decays involves the
hadronic matrix elements of vector and axial vector currents between
heavy meson states. These matrix elements can be parametrized as
\begin{eqnarray}
   \langle D_1(v',\epsilon)|\bar c\gamma^\mu b|\bar B(v)\rangle
   &=& f_{V_1}(w)\,\epsilon^{*\mu} + \Big[ f_{V_2}(w)\,v^\mu
    + f_{V_3}(w)\,v^{\prime\mu} \Big]\,\epsilon^*\cdot v \,,
    \nonumber\\
   \langle D_1(v',\epsilon)|\bar c\gamma^\mu\gamma_5\,b
   |\bar B(v)\rangle
   &=& i f_A(w)\,\varepsilon^{\mu\alpha\beta\gamma}\,
    \epsilon_\alpha^* v'_\beta v_\gamma \,, \nonumber\\
   \langle D_2^*(v',\epsilon)|\bar c\gamma^\mu\gamma_5\,b
   |\bar B(v)\rangle
   &=& k_{A_1}(w)\,\epsilon^{*\mu\alpha} v_\alpha + \Big[
    k_{A_2}(w)\,v^\mu + k_{A_3}(w)\,v^{\prime\mu} \Big]\,
    \epsilon^{*\alpha\beta} v_\alpha v_\beta \,, \nonumber\\
   \langle D_2^*(v',\epsilon)|\bar c\gamma^\mu b|\bar B(v)\rangle
   &=& i k_V(w)\,\varepsilon^{\mu\alpha\beta\gamma}\,
    \epsilon_\alpha^{*\delta} v_\delta v'_\beta v_\gamma \,,
\end{eqnarray}
where $w=v\cdot v'$, and a mass-independent normalization of meson
states is implied. The polarisation vector of the spin-one state $D_1$
satisfies $\epsilon^*\cdot v'=0$; the symmetric, traceless
Rarita-Schwinger spinor of the spin-two state $D_2^*$ obeys the
constraint $\epsilon^{*\mu\alpha} v'_\alpha=0$.

The heavy mesons can be described by spin wave-functions with
well-defined transformation properties under the Lorentz group and
heavy-quark symmetry. To leading order in $1/m_Q$, expressions for the
$\bar B\to D^{**}$ transition matrix elements are readily obtained
using the covariant trace formalism developed in Ref.~\cite{Falk}. All
form factors are proportional to a universal function $\tau_{\frac
32}(w)$, with calculable coefficients including the short-distance
corrections to the decay amplitudes \cite{review}. Explicitly, we find
\begin{eqnarray}
   \sqrt{2}\,f_{V_1}(w) &=& (w^2-1)\,C_1\,\tau_{\frac 32}(w) \,,
    \nonumber\\
   \sqrt{2}\,f_{V_2}(w) &=& \Big[ 3 C_1 + 2(w+1) C_2 \Big]\,
    \tau_{\frac 32}(w) \,, \nonumber\\
   \sqrt{2}\,f_{V_3}(w) &=& \Big[ (2-w) C_1 + 2(w+1) C_3 \Big]\,
    \tau_{\frac 32}(w) \,, \nonumber\\
   \sqrt{2}\,f_A(w) &=& (w+1)\,C_1^5\,\tau_{\frac 32}(w) \,,
    \nonumber\\
   \frac{1}{\sqrt 3}\,k_{A_1}(w) &=& (w+1)\,C_1^5\,\tau_{\frac 32}(w)
    \,, \nonumber\\
   \frac{1}{\sqrt 3}\,k_{A_2}(w) &=& - C_2^5\,\tau_{\frac 32}(w) \,,
    \nonumber\\
   \frac{1}{\sqrt 3}\,k_{A_3}(w) &=& - (C_1^5 + C_3^5)\,
    \tau_{\frac 32}(w) \,, \nonumber\\
   \frac{1}{\sqrt 3}\,k_V(w) &=& C_1\,\tau_{\frac 32}(w) \,.
\label{HQL}
\end{eqnarray}
Here $C_i$ and $C_i^5$ are the Wilson coefficients appearing in the
heavy-quark expansion of the weak currents (explicit expressions for
these functions, which depend on $w$ and the heavy-quark masses, can be
found in Ref.~\cite{review}).

It follows from these results that, in the heavy-quark limit, the
matrix elements of the weak currents vanish at the zero-recoil point
$w=1$, reflecting the fact that the ground-state $B$ meson is
orthogonal to the p-wave $D^{**}$ states. The important observation
made in Ref.~\cite{Leib} was that the leading $1/m_Q$ corrections at
zero recoil can be calculated in a model-independent way in terms of
the masses of charm-meson states. In this reference, rather complicated
formulae have been derived where the various power corrections to the
form factors are parametrized in terms of unknown, subleading universal
functions. However, all model-independent information can be
incorporated if the result for the form factor $f_{V_1}(w)$ in
(\ref{HQL}) is modified according to
\begin{eqnarray}
   \sqrt{2}\,f_{V_1}(w) &=& (w+1)(w-1+2\delta)\,C_1\,\tau_{\frac 32}(w)
    \,, \nonumber\\
   \delta &=& \left( 1 + \frac{C_3}{C_1} \right)_{w=1}\times
    \frac{M_{D_1}-M_D}{M_D}\approx 0.29 \,,
\label{deldef}
\end{eqnarray}
so that $f_{V_1}(1)$ no longer vanishes. There are many other sources
of $1/m_Q$ corrections; however, they do not yield contributions at
zero recoil. Since the physical range of $w$ values is restricted
between 1 and $\approx 1.3$, it is likely that these other corrections
will have a subdominant effect. We shall thus neglect them in our
analysis.

The differential semileptonic decay rates are given by
\begin{eqnarray}
   \frac{{\mathrm d}\Gamma(\bar B\to D_1\,\ell^-\bar\nu)}
    {{\mathrm d}w}
   &=& \frac{G_F^2\,|V_{cb}|^2 M_B^5}{48\pi^3}\,r_1^3\sqrt{w^2-1}
    \bigg\{ \left[ (w-r_1)\,f_{V_1} + (w^2-1)\,(r_1 f_{V_2}
    + f_{V_3}) \right]^2 \nonumber\\
   &&\quad\mbox{}+ 2(1-2w r_1 +r_1^2)\,\left[ f_{V_1}^2
    + (w^2-1)\,f_A^2 \right] \bigg\} \,, \nonumber\\
   \frac{{\mathrm d}\Gamma(\bar B\to D_2^*\,\ell^-\bar\nu)}
    {{\mathrm d}w}
   &=& \frac{G_F^2\,|V_{cb}|^2 M_B^5}{72\pi^3}\,r_2^3\,
    (w^2-1)^{\frac 32} \bigg\{ \!\left[ (w-r_2)\,k_{A_1} + (w^2-1)\,
    (r_2 k_{A_2} + k_{A_3}) \right]^2 \nonumber\\
   &&\quad\mbox{}+ \frac 32\,(1-2w r_2 +r_2^2)\,\left[ k_{A_1}^2
    + (w^2-1)\,k_V^2 \right] \bigg\} \,,
\label{rates}
\end{eqnarray}
where $r_1=M_{D_1}/M_B\approx 0.459$ and $r_2=M_{D_2^*}/M_B\approx
0.466$. The maximal values of $w$ in the two cases are $w_0=(1+r_1^2)/2
r_1\approx 1.32$ and $w_0=(1+r_2^2)/2 r_2\approx 1.31$, respectively.
When expressing the form factors through the universal function
$\tau_{\frac 32}(w)$ by means of the relations (\ref{HQL}) and
(\ref{deldef}), it is convenient to introduce a new function
$\tau(w)\equiv C_1\,\tau_{\frac 32}(w)$. The Wilson coefficient $C_1$
is included in this definition since only the product of $\tau_{\frac
32}(w)$ with one of the short-distance coefficients is
scheme-invariant. What remains in the expressions for the semileptonic
rates are ratios of Wilson coefficients, which are scheme-independent
and free of large logarithms. For all practical purposes, it is
sufficient to evaluate these ratios at order $\alpha_s$.

Of particular interest for the further discussion are the differential
decay rates at maximum recoil, corresponding to $q^2=0$. We find
\begin{equation}
   \frac{{\mathrm d}\Gamma(\bar B\to D^{**}\ell^-\bar\nu)}
    {{\mathrm d}w} \bigg|_{w_0}
   = \frac{G_F^2\,|V_{cb}|^2 M_B^5}{768\pi^3}\,
   \frac{(1-r_i)^5 (1+r_i)^7}{r_i^2}\,\tau^2(w_0)\,K^2(D^{**}) \,,
\end{equation}
where
\begin{eqnarray}
   K(D_1) &=& 1 + \frac{\delta}{1-r_1}
    + \frac{1+r_1}{2r_1} \left( r_1 \frac{C_2}{C_1} + \frac{C_3}{C_1}
    \right) \approx 0.95 + 1.85\delta \,, \nonumber\\
   K(D_2^*) &=& \frac{C_1^5}{C_1} - \frac{1-r_2}{2r_2} \left( r_2
    \frac{C_2^5}{C_1} + \frac{C_3^5}{C_1} \right) \approx 0.90 \,,
\end{eqnarray}
and the short-distance coefficients are evaluated at $w_0$.
Numerically, we obtain
\begin{eqnarray}
   \frac{{\mathrm d}\Gamma(\bar B\to D_1\,\ell^-\bar\nu)}
    {{\mathrm d}w} \bigg|_{w_0}
   &\approx& 0.160\,(1 + 1.944\delta)^2\times [\tau(1.32)]^2\,
    \mbox{ps}^{-1} \nonumber\\
   &\approx& 0.394\times [\tau(1.32)]^2\,\mbox{ps}^{-1} \,, \nonumber\\
   \frac{{\mathrm d}\Gamma(\bar B\to D_2^*\,\ell^-\bar\nu)}
    {{\mathrm d}w} \bigg|_{w_0}
   &\approx& 0.136\times [\tau(1.31)]^2\,\mbox{ps}^{-1} \,.
\label{Gamsl}
\end{eqnarray}

To calculate the total semileptonic rates requires an ansatz for the
form factor $\tau(w)$. Since the accessible range of $w$ values is
small, it is sufficient to adopt a linear approximation with a slope
parameter $\rho^2$: $\tau(w) = \tau(1)\,[ 1 - \rho^2(w-1) + \dots ]$.
Higher-order terms can be partially taken into account by reexpressing
the results for the decay rates obtained in linear approximation
through values of the function $\tau(w)$ at intermediate points. In
that way, we find
\begin{eqnarray}
   \Gamma(\bar B\to D_1\,\ell^-\bar\nu)
   &\approx& (0.0158 + 0.0694\delta + 0.1169\delta^2)
    \times [\tau(1.23)]^2\,\mbox{ps}^{-1} \nonumber\\
   &\approx& 0.0462\times [\tau(1.23)]^2\,\mbox{ps}^{-1} \,,
    \nonumber\\
   \Gamma(\bar B\to D_2^*\,\ell^-\bar\nu)
   &\approx& 0.0207\times [\tau(1.21)]^2\,\mbox{ps}^{-1} \,.
\label{Gamtot}
\end{eqnarray}

The results for the differential and total semileptonic rates in
(\ref{Gamsl}) and (\ref{Gamtot}) do not include $1/m_Q$ corrections
except those proportional to the quantity $\delta$, which are
kinematically enhanced and specific for $\bar B\to D_1$ transitions.
{}From the analysis of power corrections for $\bar B\to
D^{(*)}\ell^-\bar\nu$ decays it is known that the remaining $1/m_Q$
corrections tend to be spin independent and thus cancel, to a large
extent, in ratios of decay rates \cite{review}. Therefore, we expect
that ratios of the decay rates estimated above are accurate to about
20\%.

\section{Implications and conclusions}

We are now in a position to compare the theoretical predictions derived
above with the available experimental data. Consider first the
branching ratio for the decay $B^-\to D_1^0\,\ell^-\bar\nu$.
Multiplying the theoretical prediction for the total semileptonic rate
in (\ref{Gamtot}) by the lifetime $\tau_{B^-}=1.65$\,ps, we find
${\mathrm B}(B^-\to D_1^0\,\ell^-\bar\nu) \approx 7.9\%\times
[\tau(1.23)]^2$. This value is about 3 times larger than the result
obtained in the strict heavy-quark limit, where the kinematically
enhanced terms involving $\delta$ are neglected. Under the assumption
that ${\mathrm B}(D_1^0\to D^{*+}\pi^-)=2/3$, the experimental value in
(\ref{CLdat2}) implies ${\mathrm B}(B^-\to
D_1^0\,\ell^-\bar\nu)=(0.56\pm 0.15)\%$. Comparing this with the
theoretical result, we find that $\tau(1.23)=0.27\pm 0.04$ is required
in order to fit the data. This value is in good agreement with
theoretical predictions obtained using relativistic quark models
($\tau(1.23)\approx 0.3$ \cite{More}) or QCD sum rules
($\tau(1.23)\approx 0.22$ \cite{Cola}). We take this agreement as an
indication that the terms involving the quantity $\delta$ in
(\ref{Gamsl}) and (\ref{Gamtot}) do indeed capture the dominant
corrections to the heavy-quark limit.

Our next goal is to understand the ratio of the hadronic and
semileptonic widths for $\bar B\to D_1$ transitions. The experimental
data in (\ref{CLdat1}) and (\ref{CLdat2}) imply
\begin{equation}
   R_1 = \frac{\Gamma(B^-\to D_1^0\,\pi^-)}
    {\Gamma(B^-\to D_1^0\,\ell^-\bar\nu)} = 0.21\pm 0.08 \,,
\end{equation}
independently of the $D_1\to D^*\pi$ branching ratios. Combining
(\ref{cl3ratio}), (\ref{Gamsl}) and (\ref{Gamtot}), we obtain the
theoretical prediction (using ${\mathrm d}q^2=2 M_B M_{D_1}\, {\mathrm
d}w$)
\begin{eqnarray}
   R_1 &\approx& \frac{3\pi^2 f_\pi^2\,|V_{ud}|^2}{M_B M_{D_1}}\,
    |1+\delta_8(m_b)|^2\,
    \frac{{\mathrm d}\Gamma(B^-\to D_1^0\,\ell^-\bar\nu)/
     {\mathrm d}w \Big|_{w_0}}
     {\Gamma(B^-\to D_1^0\,\ell^-\bar\nu)} \nonumber\\
   &\approx& 0.32 \left( |1+\delta_8(m_b)|\,
    \frac{\tau(1.32)}{\tau(1.23)} \right)^2 \,,
\end{eqnarray}
which is rather insensitive to the value of $\delta$. Obviously, there
is no problem to account for the data. Making the conservative
assumption $\rho^2=1.5\pm 0.5$ for the slope parameter of the function
$\tau(w)$, we find $R_1\approx (0.20\pm 0.06)\,|1+\delta_8(m_b)|^2$,
which is in good agreement with experiment provided the nonfactorizable
corrections parametrized by $\delta_8$ are moderate in size. Solving
for these corrections we obtain $|1+\delta_8(m_b)|=1.02\pm 0.25$, in
accordance with the fact that $\delta_8=O(1/N_c)$.

Consider next the ratio of the two semileptonic rates for $B$ decays
into $D_1$ and $D_2^*$ mesons. The available experimental data for this
ratio depend on some $D^{**}\to D^*\pi$ branching ratios that have not
yet been measured directly. We define
\begin{equation}
   h = \frac{{\mathrm B}(D_1^0\to D^{*+}\pi^-)}
    {{\mathrm B}(D_2^{*0}\to D^{*+}\pi^-)} = 3.3\pm 0.8 \,,
\label{hval}
\end{equation}
where the quoted numerical value is obtained under the assumptions
stated in the introduction. The CLEO data in (\ref{CLdat2}) imply
\begin{equation}
   R_2 = \frac{\Gamma(B^-\to D_2^{*0}\,\ell^-\bar\nu)}
    {\Gamma(B^-\to D_1^0\,\ell^-\bar\nu)} = (0.16\pm 0.19) h
   = 0.53\pm 0.63 \,,
\end{equation}
or $R_2<1.48~(90\%~{\mathrm CL})$. From (\ref{Gamtot}), we obtain the
theoretical prediction
\begin{equation}
   R_2 \approx \frac{1.31}{1 + 4.39\delta + 7.40\delta^2}\,
   \left( \frac{\tau(1.21)}{\tau(1.23)} \right)^2 \approx 0.48 \,,
\end{equation}
which is in good agreement with the data. We have again assumed
$\rho^2=1.5\pm 0.5$ to estimate the form-factor difference, which is a
small effect in the present case. Although the experimental errors in
the value of $R_2$ are large, to reproduce the data requires the
presence of the terms involving the quantity $\delta$ \cite{Leib}.

Consider finally the ratio of the two hadronic rates for $B$ decays
into $D_1$ and $D_2^*$ mesons. The CLEO data in (\ref{CLdat1}) imply
\begin{equation}
   R_3 = \frac{\Gamma(B^-\to D_2^{*0}\,\pi^-)}
    {\Gamma(B^-\to D_1^0\,\pi^-)} = (0.54\pm 0.26) h = 1.8\pm 1.0 \,.
\label{R3val}
\end{equation}
Combining (\ref{cl3ratio}) with the expressions for the semileptonic
rates at maximum recoil given in (\ref{Gamsl}), and assuming again
$\rho^2\approx 1.5$ to estimate the tiny form-factor difference between
the two cases, we obtain the theoretical prediction
\begin{equation}
   R_3 \approx \frac{0.86}{(1 + 1.94\delta)^2}\,\left|
   \frac{1+\delta_8^{(D_2^*)}(m_b)}{1+\delta_8^{(D_1)}(m_b)} \right|^2
   \approx 0.35\,\left| \frac{1+\delta_8^{(D_2^*)}(m_b)}
    {1+\delta_8^{(D_1)}(m_b)} \right|^2 \,,
\label{R3th}
\end{equation}
which is significantly lower than the data. It must be stressed,
however, that the experimental errors are large, and the discrepancy
between theory and experiment is only about 1.5 standard deviations.
Indeed, it would be very difficult to reconcile the central
experimental value in (\ref{R3val}) with the theoretical expectation.
The least understood aspect of the theoretical calculation is the
question about the size of the nonfactorizable contributions to the
hadronic decay widths. However, to reproduce the data would require
that
\begin{equation}
   \left| \frac{1+\delta_8^{(D_2^*)}(m_b)}{1+\delta_8^{(D_1)}(m_b)}
   \right| \approx 2.27\pm 0.63 \,.
\end{equation}
Since the deviation of this ratio from unity is of order $1/N_c$ in the
large-$N_c$ limit, such a large value seems very unlikely. This is even
more so as we have shown above that the nonfactorizable contributions
to the $B^-\to D_1^0\,\pi^-$ decay rate are very small. Another
possibility would be to blame the discrepancy between (\ref{R3val}) and
(\ref{R3th}) on the $D^{**}\to D^*\pi$ branching ratios, which have not
yet been measured directly. To reproduce the data would require that
$h\approx 0.65\pm 0.31$, which is much smaller than the commonly
assumed value quoted in (\ref{hval}). Besides the fact that such a
small value would imply exotic (or at least not well understood)
physics in the strong decays of p-wave charm mesons, this possibility
is essentially ruled out by the good agreement of the theoretical
prediction for the ratio $R_2$ with the experimental value for that
ratio derived assuming the standard value for $h$. Finally, one may ask
whether the theoretical prediction (\ref{R3th}) could be spoiled by
$1/m_Q$ corrections not included in our estimate of the semileptonic
rates in (\ref{Gamsl}). However, as we have argued above, those
corrections tend to cancel in ratios of decay rates and are expected
not to exceed the level of 20\%.

To summarize, whereas there is in general good agreement between
theoretical expectations and experimental data for semileptonic and
hadronic $B$ decays into final states containing a p-wave charm meson,
the experimental central value of the branching ratio for the decay
$B^-\to D_2^{*0}\,\pi^-$ cannot be accommodated by theory. We predict
that future, more accurate measurements of this branching ratio will
find a value
\begin{equation}
   {\mathrm B}(B^-\to D_2^{*0}\,\pi^-) \approx
   0.35\,{\mathrm B}(B^-\to D_1^0\,\pi^-) \approx 4\times 10^{-4} \,,
\end{equation}
which is about a factor 5 lower than the current central value reported
by the CLEO Collaboration \cite{CLEOhad}. A similar conclusion has been
reached recently in Ref.~\cite{Muno} from a calculation based on the
non-relativistic quark model. If, on the other hand, the current
central value were confirmed, this would pose a serious problem for the
theory of $B$ decays.

\vspace{0.5cm}
{\it Acknowledgements:\/}
We are grateful to J. Bjorken, M. Beneke and B.~Stech for helpful
discussions.


\newpage
\begin{thebibliography}{99}

\bibitem {CLEOhad}
CLEO Collaboration (J. Gronberg et al.), Conference report CLEO CONF
96-25 (1996), contributed paper to the 28th International Conference on
High Energy Physics (ICHEP 96), Warsaw, Poland, July 1996.

\bibitem {CLEOsl}
CLEO Collaboration (A. Anastassov et al.), Preprint CLNS 97/1501
(1997), contributed paper to the 1997 International Europhysics
Conference on High Energy Physics (HEP 97), Jerusalem, Israel, August
1997.

\bibitem {ALEPHsl}
ALEPH Collaboration (D. Buskulic et al.), Z.\ Phys.\ C {\bf 73}, 601
(1997).

\bibitem {CLE}
CLEO Collaboration (P. Avery et al.), Phys.\ Lett.\ B {\bf 331}, 236
(1994).

\bibitem {ARG}
ARGUS Collaboration (H. Albrecht et al.), Phys.\ Lett.\ B {\bf 232},
398 (1989).

\bibitem {Bj89}
J.D. Bjorken, Nucl.\ Phys.\ B (Proc.\ Suppl.) {\bf 11}, 325 (1989).

\bibitem {BSW}
M. Bauer, B. Stech and M. Wirbel, Z.\ Phys.\ C {\bf 34}, 103 (1987).

\bibitem {NS}
M. Neubert and B. Stech, Preprint CERN-TH/97-99 [hep-ph/9705292], to
appear in {\em Heavy Flavours}, Second Edition, edited by A.J.~Buras
and M.~Lindner (World Scientific, Singapore);\\
M. Neubert, Preprint CERN-TH/97-169 [hep-ph/9707368], to appear in the
Proceedings of the High-Energy Physics Euroconference on Quantum
Chromodynamics (QCD 97), Montpellier, France, 3--9 July 1997.

\bibitem {Chen}
H.Y. Cheng, Phys.\ Lett.\ B {\bf 335}, 428 (1994).

\bibitem {Soar}
J.M. Soares, Phys.\ Rev.\ D {\bf 51}, 3518 (1995).

\bibitem {Falk}
A.F. Falk, H. Georgi, B. Grinstein and M.B. Wise, Nucl.\ Phys.\ B
{\bf 343}, 1 (1990);\\
A.F. Falk, Nucl.\ Phys.\ B {\bf 378}, 79 (1992).

\bibitem {review}
M. Neubert, Phys.\ Rep.\ {\bf 245}, 259 (1994).

\bibitem {Leib}
A.K. Leibovich, Z. Ligeti, I.W. Steward and M.B. Wise, Phys.\ Rev.\
Lett.\ {\bf 78}, 3995 (1997); Preprint CALT-68-2120 (1997)
[hep-ph/9705467].

\bibitem {More}
V. Mor\'enas et al., Preprint LPTHE Orsay-97/19 (1997)
[hep-ph/9706265].

\bibitem {Cola}
P. Colangelo, G. Nardulli and N. Paver, Phys.\ Lett.\ B {\bf 293}, 207
(1992).

\bibitem {Muno}
G. Lopez Castro and J.H. Munoz, Phys.\ Rev.\ D {\bf 55}, 5581 (1997).

\end{thebibliography}
\end{document}